\documentclass[sigconf]{acmart}

\usepackage{paralist}
\usepackage{indentfirst}
\usepackage{subfigure}
\usepackage{multirow}

\AtBeginDocument{%
  \providecommand\BibTeX{{%
    \normalfont B\kern-0.5em{\scshape i\kern-0.25em b}\kern-0.8em\TeX}}}

\begin{document}
\copyrightyear{2022}
\acmYear{2022}
\setcopyright{acmcopyright}\acmConference[ICPC '22]{30th International Conference on Program Comprehension}{May 16--17, 2022}{Virtual Event, USA}
\acmBooktitle{30th International Conference on Program Comprehension (ICPC '22), May 16--17, 2022, Virtual Event, USA}
\acmPrice{15.00}
\acmDOI{10.1145/3524610.3527915}
\acmISBN{978-1-4503-9298-3/22/05}

\title{Unified Abstract Syntax Tree Representation Learning for Cross-Language Program Classification}

% \author{Kesu Wang$^1$, Meng Yan$^2$, He Zhang$^1$, Haibo Hu$^2$}
% \affiliation{%
%  \institution{$^1$State Key Lab for Novel Software Technology, Nanjing University, Nanjing, China}
%  \institution{$^2$School of Big Data \& Software Engineering, Chongqing University, Chongqing, China}
% }

\author{Kesu Wang$^1$, Meng Yan$^2$, He Zhang$^{1*}$, Haibo Hu$^2$}
% \author{He Zhang$^1$}
% \authornote{corresponding author}

% 方案1：
\affiliation{%
 \institution{$^1$State Key Lab for Novel Software Technology, Nanjing University, Nanjing\country{China}; $^*$Corresponding author}
 \institution{$^2$School of Big Data \& Software Engineering, Chongqing University, Chongqing\country{China}}
%  \institution{$^*$Corresponding author}
}
\email{kesucaso@163.com, mengy@cqu.edu.cn, hezhang@nju.edu.cn, haibo.hu@cqu.edu.cn}

% 方案2：
% \author{Kesu Wang}
% \affiliation{\institution{State Key Lab for Novel Software Technology, \\Nanjing University, Nanjing, China}}

% \author{Meng Yan}
% \affiliation{\institution{School of Big Data \& Software Engineering, \\Chongqing University, Chongqing, China}}

% \author{He Zhang}
% \affiliation{\institution{State Key Lab for Novel Software Technology, \\Nanjing University, Nanjing, China}}
% \authornote{Corresponding author}

% \author{Haibo Hu}
% \affiliation{\institution{School of Big Data \& Software Engineering, \\Chongqing University, Chongqing, China}}

\begin{abstract}
Program classification can be regarded as a high-level abstraction of code, laying a foundation for various tasks related to source code comprehension, and has a very wide range of applications in the field of software engineering, such as code clone detection, code smell classification, defects classification, etc. The cross-language program classification can realize code transfer in different programming languages, and can also promote cross-language code reuse, thereby helping developers to write code quickly and reduce the development time of code transfer. Most of the existing studies focus on the semantic learning of the code, whilst few studies are devoted to cross-language tasks. The main challenge of cross-language program classification is how to extract semantic features of different programming languages. In order to cope with this difficulty, we propose a \underline{U}nified \underline{A}bstract \underline{S}yntax \underline{T}ree (namely UAST in this paper) neural network. In detail, the core idea of UAST consists of two unified mechanisms. First, UAST learns an AST representation by unifying the AST traversal sequence and graph-like AST structure for capturing semantic code features. Second, we construct a mechanism called unified vocabulary, which can reduce the feature gap between different programming languages, so it can achieve the role of cross-language program classification. Besides, we collect a dataset containing 20,000 files of five programming languages, which can be used as a benchmark dataset for the cross-language program classification task. We have done experiments on two datasets, and the results show that our proposed approach outperforms the state-of-the-art baselines in terms of four evaluation metrics (Precision, Recall, F1-score, and Accuracy).
\end{abstract}

% \keywords{program classification, cross-language program classification, code learning, code embedding, neural network}

% Graph Neural Network

\maketitle

\section{Introduction}
Program classification is aiming to automatically classify programs according to their functions or semantics. As one of the primary means of facilitating program comprehension, program classification nowadays has been widely employed in a variety of different tasks such as code clone detection~\cite{baker1993program, bellon2007comparison, borstler1995feature}, defect detection and identification~\cite{wang2016automatically, li2017software}, and code search~\cite{kim2018facoy}, etc. The benefits of program classification have been generally recognized and valued, program classification has long relied on manual classification, which is very time-consuming and error prone. In recent years, more and more researchers have paid attention to source code-oriented program comprehension, as source codes are the most natural representation of programs. Moreover, source codes are well-structured, are ideal to support automating program comprehension.

Cross-language program classification refers to how programs written in different programming languages can be classified by their functions according to the structure and semantics of their codes. For example, although the code structure and logic of \textit{"quick sort" } written in C++ and \textit{"bubble sort"} written in Python is  implemented differently, they are both sorting algorithms in essence, so they should be classified as codes of the same function. Consequently, cross-language program classification can reduce the time to implement programs of the same function in different languages, and can also promote cross-language code reuse, thereby helping developers write code quickly and reduce development time for code transfer.

However, it is challenging to capture code semantics efficiently and accurately. Moreover, considering that different programming languages have different grammatical rules and coding features, it is even more difficult to accurately extract cross-language code semantics. In terms of the extraction of code semantic information, Li et al.~\cite{li2017cclearner} and Harer et al.~\cite{harer2018automated} use tokens to generate embedded vectors and feed them into neural networks for code feature learning. However, a token only contains lexical information of the code and cannot reflect the structural and semantic characteristics of the code. Ben et al.~\cite{ben2018neural} use Intermediate Representations (IRs) to design a graph called XFG, and then uses neural networks (GNN~\cite{scarselli2008graph}, RNN~\cite{elman1990finding}) to learn the semantic features of the graph. Since the XFG contains data dependence of the code, it is helpful for semantic extraction, but obtaining IR requires code compilation, which makes it impossible to process some incomplete code fragments. Azcona et al.~\cite{azcona2019user2code2vec} and Mou et al.~\cite{mou2016convolutional} propose code learning approaches based on Abstract Syntax Tree (AST), which could learn the features from the traversal or the tree structure of the AST, and have a certain semantic effect.

\begin{figure*}[t]
% \begin{figure*}[!htbp]
    \centering
    \vspace*{-2.0ex}
    \includegraphics[width=\textwidth]{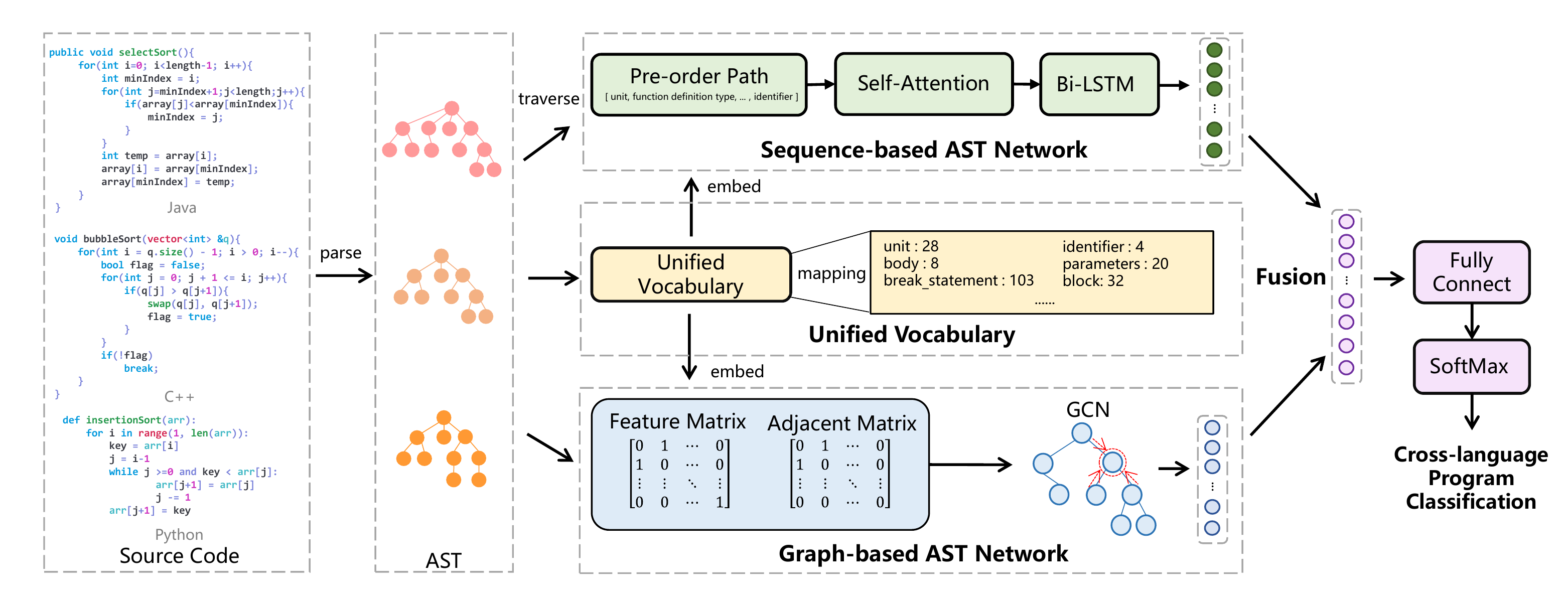}
    \vspace*{-2.0ex}
    \caption{An overview of the proposed approach -- UAST.}
    \label{fig:framework}
    \vspace*{-2.0ex}
\end{figure*}

These researches for learning the semantic features of the code mentioned above are all set for single language, and cannot learn the semantic features of different programming languages, resulting in not much breakthrough in the research of cross-language program classification. Bui et al.~\cite{bui2019bilateral} pay attention to cross-language issues, and propose a Bi-NN framework to learn the semantic features of two different programming languages. Specifically, it uses two networks of the same structure (similar to Siamese \cite{chopra2005learning} networks) to perform code feature learning. Although the model can handle cross-language program classification problems, it is essentially learning a single programming language feature respectively and fusing different learned code features. In addition, the structure is not highly scalable, if the number of programming languages increases, a new network needs to be added, which makes the network structure more complicated and more time-consuming on the training time. Bui et al.~\cite{bui2021infercode} propose a pre-trained model with the idea of self-supervision. It trains code features on the large corpus which contains multiple programming languages, so it can be used for cross-language program classification tasks. However, it does not have special preprocessing for different languages and just directly generates AST for mixed training, which essentially does not consider the differences of different languages.

To address those problems mentioned above, we propose a neural network called \textbf{U}nified \textbf{A}bstract \textbf{S}yntax \textbf{T}ree (\textbf{UAST} for short). With regard to semantic extraction, we employ self-attention combined with Bi-LSTM to extract flattened AST sequence features, which can capture the global logical structure characteristics of the code. Besides, we use Graph Convolutional Neural (GCN) \cite{kipf2016semi} network to extract the local feature of the graph-like AST structure. And then, we fuse those two features (sequence features and graph features) to strengthen the feature of corresponding dimension, so that the structural and semantic characteristics of the code can be obtained. For cross-language semantic learning, we have established a unified vocabulary for embedded mapping and use this vocabulary to reduce the differences between different programming languages, so as to facilitate the learning of neural networks. In addition, we collect a dataset that contains five different programming languages from Leetcode\footnote{https://leetcode.com/}. We conduct experiments on two datasets, the results both show that our UAST performs better than the state-of-the-art baselines by 4.54\% - 22.62\% in terms of four evaluation metrics. Furthermore, we conduct ablation experiments to explore the impact of our unified vocabulary and unified AST feature fusion on model performance. 

In summary, the main contributions of this study can be summarized as follows.
\begin{itemize}
    \item We propose a unified AST representation learning approach using two sub-networks (SAST and GAST). SAST is used to extract the global code syntactic features contained in the AST path sequence. GAST is used to capture the local code semantic features in the AST tree. The unified AST network can comprehensively consider global and local learned code features, which could effectively learn code semantic features.
    \item We construct a unified vocabulary mechanism to reduce the difference between different programming languages. The initial input embedded vector can be obtained through the vocabulary mapping, and then put the vector to two sub-networks for training, which can classify cross-language programs.
    \item We conduct experiments on two datasets, and the experimental results show that our performance is better than other state-of-the-art baselines (CodeBERT, Infercode) in terms of Recall, Precision, F1-score and Accuracy.
    \item We contribute a benchmark dataset for the cross-language program classification task. The dataset contains five programming languages (C, C++, Java, Python, and JavaScript), with a total of 50 problems and 20,000 solution files. The solutions to each problem are semantically similar codes to each other.\footnote{Our replication package including both datasets and scripts can be found at https://github.com/kkcookies99/UAST.}
\end{itemize}

% The rest of this paper is organized as follows. Section 2 describes the proposed network UAST in details. Section 3 and Section 4 discuss the experiment design and results.  Section 5 presents the related work. Section 6 demonstrates the discussion. Finally, Section 7 concludes the work of this paper and presents the future work.

The remainder of this paper is organized as follows. Section~\ref{SEC:APP} elaborates on the proposed approach. Sections~\ref{SEC:EXP} and~\ref{SEC:RES} report on the experimental design and results respectively.  Section~\ref{SEC:DIS} discusses the considerations behind the proposed approach. Section~\ref{SEC:REL} reviews the related work. Finally, we conclude this paper and present the future work in Section~\ref{SEC:CON}.

\section{The Proposed Approach}
\label{SEC:APP}

In this section, we present the design and implementation details of our proposed \textbf{UAST} (Unified Abstract Syntax Tree) neural network model for cross-language program classification.

\begin{figure*}[t]
\centering
\includegraphics[width=\textwidth]{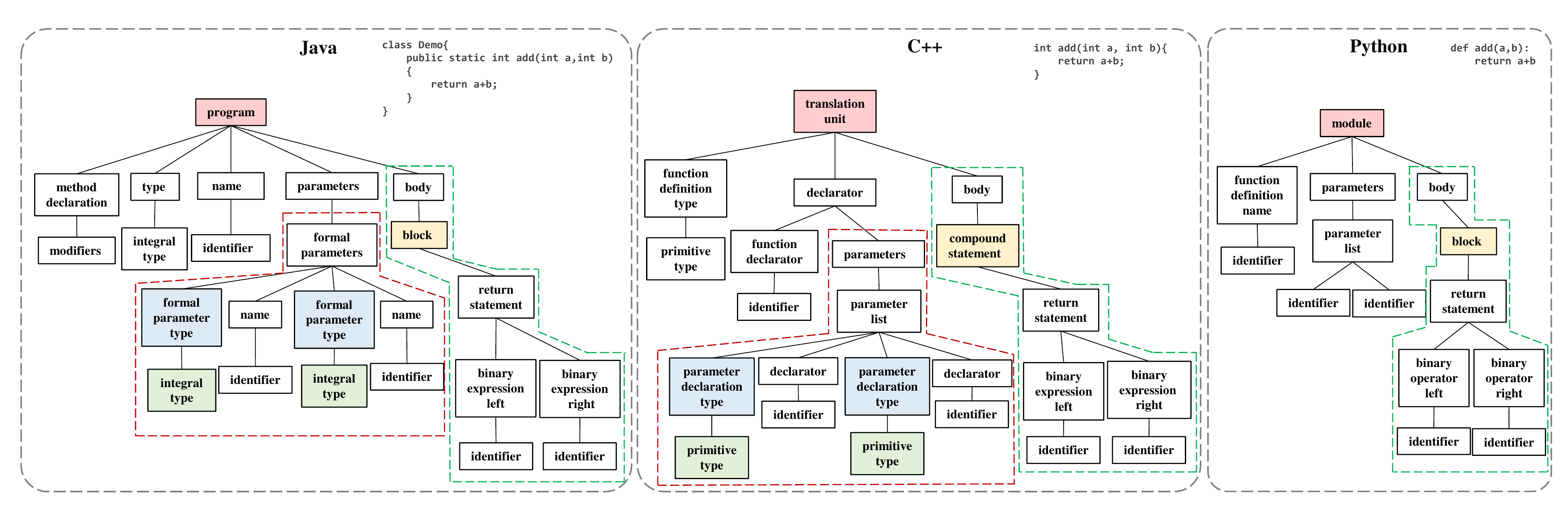}
\caption{The construction of unified vocabulary. Boxes of the same color have the same meaning, and the areas surrounded by dotted lines of the same color represent the same code structure.}
\label{fig:uv}
\end{figure*}

\subsection{Overall Structure}
As illustrated in Figure \ref{fig:framework}, this model firstly takes different programming source codes as input and then parses them into ASTs. Afterwards, it performs path embedding and graph embedding using unified vocabulary on ASTs. Then the path embedded vector and the graph embedded vector are fed into different sub-networks to capture code features respectively. Finally, it fuses separate learned features and conducts the classification task.

Specifically, it mainly consists of five parts: Unified Vocabulary, Sequence-based AST Network, Graph-based AST Network, Unified AST Feature Fusion, and Cross-language Program Classification. We explain the details of each part in the following subsections.

\subsection{Unified Vocabulary}
Since different programming languages have different coding rules and coding characteristics, there are certain differences in the text representation of different languages. Figure \ref{fig:uv} shows ASTs generated by tree-sitter\footnote{http://tree-sitter.github.io/tree-sitter/} parsing the function of adding two integers implemented by three programming languages (Java/C++/Python). The reason why we choose tree-sitter for syntactic analysis of the source code is that it supports 40 programming languages, and also provides a Python API which is easy to use. It can be found in Figure \ref{fig:uv} that there are some differences in the node names of ASTs. The root node is called \textit{"program"} in the Java AST, \textit{"translation unit"} in C++ AST, and \textit{"module"} in Python AST. Actually, these three terminologies all represent the same semantics, that is, the coding unit, which generally means the code file or the program. In order to reduce the difference in node names between different languages, we propose a mechanism called "Unified Vocabulary", which is mainly used to normalize AST node names in different languages, such as the "coding unit" mentioned above is unified into "unit". Specifically, we use \textit{"unit" }to replace \textit{"program, translation unit, module"} uniformly, which will reduce the difference between different coding languages. Furthermore, \textit{"block"} in Java, "\textit{compound statement}" in C++ and "\textit{block}" in Python are essentially a code block, so \textit{"block"} is used instead. The nodes of the same color in Figure \ref{fig:uv} represent the same meaning, so they all will be processed into the same node name uniformly.

We have considered all similar but different expressions of node names in different languages. The unified vocabulary mechanism will unify AST node names generated by different programming languages, alleviating the differentiation caused by different coding characteristics. Thus, the embedded vector generated using the unified vocabulary can learn the feature of various programming languages and thereby tackle the problem of cross-language programming classification.

\subsection{Sequence-based AST Network}
The foundation of cross-language program classification is to learn semantic and syntactic code features of different programming languages. As for the extraction of the syntactic structure information of the code, we propose a \textbf{S}equence-based \textbf{AST} network (\textbf{SAST} for short). First, we perform a pre-order traversal of the unified AST, and the obtained path sequence can be regarded as a flattened presentation of the AST. The path sequence contains the global information of the source code, and also shows the syntactic structure characteristics of the source code to a certain extent. In order to extract the dependencies between the nodes within the sequence, we use the self-attention structure \cite{vaswani2017attention}. The self-attention mechanism is a powerful mechanism in the transformer structure, which is very effective in extracting internal relationships and can alleviate the problem of long-distance dependence. The calculation formula is as follows:

\begin{equation}
    Attention(Q,K,V)=SoftMax(\frac{QK^T}{\sqrt{d_k}})V
\end{equation}
where the three matrices $Q\in \mathbb{R}^{l\times d}$, $K\in \mathbb{R}^{l\times d}$, and $V\in \mathbb{R}^{l\times d}$ are initialized and generated according to the embedded path sequence vector, and these three matrices are equal in the self-attention mechanism, which could reduce the parameters of the model and can train faster. $d$ is the embedding dimension of the path sequence. $l$ is the length of input path. Dot product is calculated between $Q$ and $K$. $d_k \in \mathbb{R}^d$ is the dimension of input vector.  And $Attention(Q,K,V)$ is the calculated attention score.

In writing code, the context of the code statement often reflects its intent. For example, it needs to declare a variable before using it in C++. The above-mentioned self-attention mechanism has captured the internal relationship of the code embedded vector. So in addition to extracting the internal dependencies of the input source code, it also needs to capture the context dependency of the source code. Therefore, the Bidirectional Long Short-Term Memory (Bi-LSTM) \cite{graves2005framewise} is introduced here. The Bi-LSTM could learn features of the input data from two directions, so it can infer the current information from the context of the code. In our proposed SAST network, Bi-LSTM is used to comprehensively consider all available input path information in the context to extract semantic and logic features of the source code. Specifically, the hidden state of the LSTM at each position $t$ of the input path is computed as:

\begin{gather}
    i_t=\sigma(W_i\cdot[h_{t-1},x_t]+b_i)\\
    f_t=\sigma(W_f\cdot[h_{t-1},x_t]+b_f)\\
    o_t=\sigma(W_o\cdot[h_{t-1},x_t]+b_o)\\
    \tilde{c}_t=tanh(W_C\cdot[h_{t-1},x_t]+b_c)\\
    c_t=f_t\odot c_{t-1}+i_t\odot \tilde{c}_t\\
    h_t=o_t\odot tanh(c_t)
\end{gather}
where $\sigma$ is the sigmoid function, $tanh$ is the hyperbolic function, $x_t \in\mathbb{R}^d $ presents the data at position $t$ of the input path sequence after self-attention, $c_t$ presents the hidden unit state of $x_t$, and $h_t \in \mathbb{R}^h$ represents the hidden unit state of the learning layer, which is the final extracted code features. $W_i, W_f, W_o \in \mathbb{R}^{h\times d}$ are the trainable weight matrices. $\odot$ is the element-wise matrix multiplication operator.

\vspace{-0.6cm}
\begin{equation}
    {h_{SAST}} = \mathop{h_t}\limits ^{\rightarrow} \oplus \mathop{h_t}\limits ^{\leftarrow}
\end{equation}

After that, we concatenate the hidden state $\mathop{h_t}\limits ^{\rightarrow}\in \mathbb{R}^{h}$ learned by the forward LSTM and the hidden state $\mathop{h_t}\limits ^{\leftarrow}\in \mathbb{R}^h$ learned by the backward LSTM to obtain the ${h_{SAST}}\in \mathbb{R}^{2\times h}$, which contains the context features of the code.

%summary
% Our proposed SAST network is used to learn the 

\subsection{Graph-based AST Network}
Sequence-based AST network has learned the global structure and syntactic features of the code from the path sequence, but the path sequence is a flattened representation, which leads to ignoring some tree-like structure information of the code. We find that except for language-specific library files or package files, the logic of different programming languages is generally the same when writing specific functions. As shown in the Figure \ref{fig:uv}, the parts enclosed by the dashed lines of the same color all have the same meaning (i.e., the parts outlined by the green dotted line all represent \textit{"return a+b"} statement in 3 programming languages). In addition, the parts enclosed by the dashed lines of the same color are also highly consistent in the AST shape. In order to learn the local tree features of the AST, it is easy to think of using GCN to extract the local information of the code, so we propose the \textbf{G}raph-based \textbf{AST} network (\textbf{GAST} for short). We regard the AST structure as a special graph and perform convolution operation on it to extract code semantic features. GCN can aggregate and learn features of neighbor nodes of each node in the unified AST, so that it can capture the local features of the AST, that is, the local semantics of the code.

\begin{figure}[h]
\centering
\includegraphics[width=\linewidth]{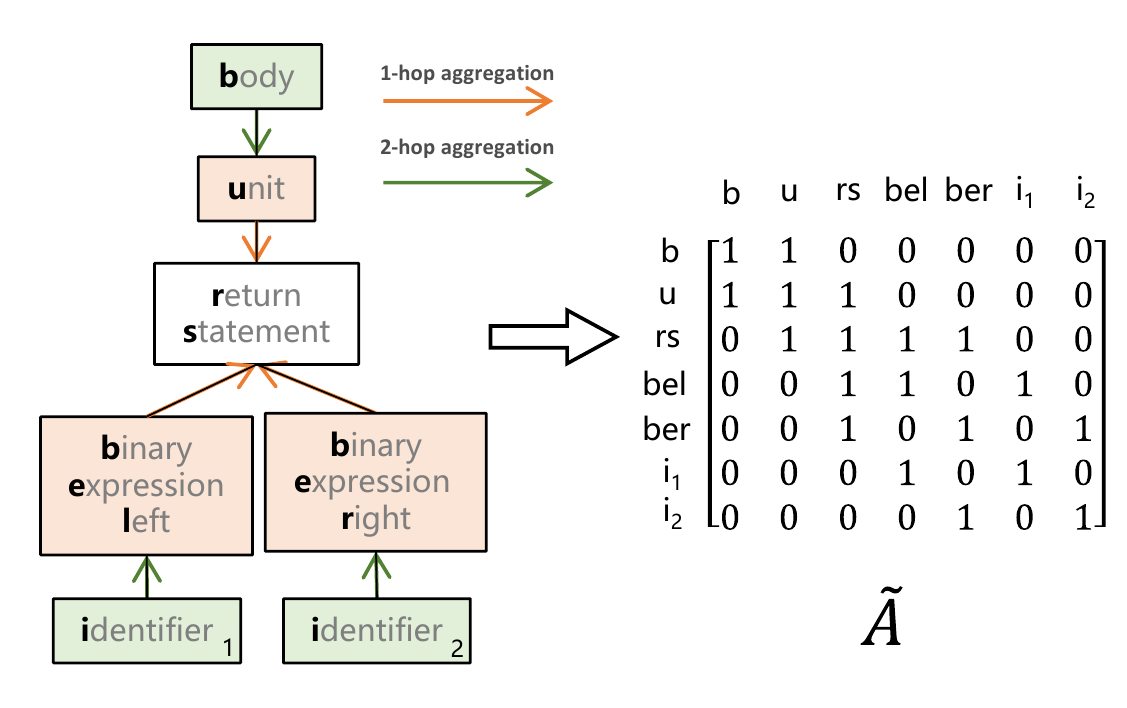}
\caption{A case of AST node aggregation of GAST.}
\label{fig:GAST}
\end{figure}

Figure~\ref{fig:GAST} illustrates the AST of \textit{"return a+b"} statement and a two-layer convolution operation on it. Specifically, the \textit{"return statement"} node will aggregate the information of its first-order neighbors (\textit{"unit, binary expression left, and binary expression right"}) with the 1-hop aggregation, which can extract the direct relationship between code statements. As the number of hops increases, the \textit{"return statement"} node can also indirectly aggregate the information of their second-order neighbors(\textit{"body, $identifier_1$, $identifier_2$"}). Therefore, the \textit{"return statement"} node could learn the local structural and semantic features of its neighbors. The right of Figure \ref{fig:GAST} is the adjacency matrix of the left AST, and it is worth noting that $ \widetilde{A}$ adds an identity matrix to its adjacency matrix, indicating that the node could also learn the feature of itself. The graph convolution operation is defined as follows:

% Due to the general local code structure, we set its k-hop to 2 when performing the experiment, stacking two layers of GCN layers to let the node learn the information of its 2-hop neighbors. On the right of Figure \ref{fig:GAST} is the adjacency matrix of the left AST with the identity matrix, so it can combine the feature of itself.

\begin{gather}
    \widetilde{A} = A + I \\
    {H_i}^{(l+1)}=\sigma(\sum_{j\in N} D^{-\frac{1}{2}} \widetilde{A} D^{-\frac{1}{2}} {H_j}^{(l)}W^{(l)})
\end{gather}
where ${H_i}^{(l+1)}$ is the feature of node $i$  in the layer ${(l+1)}$, ${H_j}^{(l)}$ is the feature of all neighbor nodes of node $i$ (including itself) in the layer $l$; $N$ is the number of all neighbors of node $i$; $\widetilde{A}\in\mathbb{R}^{N\times N}$ is the adjacency matrix $A$ of node $i$ added with the identity matrix; $D\in\mathbb{R}^{N\times N}$ is the degree matrix of $\widetilde{A}$;  $W^{(l)}\in \mathbb{R}^{d_{in}\times d_{out}}$ is the trainable weight matrix in the layer $l$.

\subsection{Unified AST Feature Fusion}
The aforementioned SAST has captured the global structure and logical characteristics of the code, and GAST has extracted the local structural and semantic feature of the code. In order to comprehensively consider the global and local code semantic feature, a fusion mechanism is further needed. We realize the enhancement of dimensional features through vector concatenation, which can be described as:
\begin{gather}
    h_{code}=concat(h_{SAST},h_{GAST})
\end{gather}
where $h_{code}\in\mathbb{R}^{2\times h+d_{out}}$ represents the feature vector after the unified AST feature fusion, $h_{SAST}\in\mathbb{R}^{2\times h}$ represents the global structural feature learned from the flattened sequence, and $h_{GAST}\in\mathbb{R}^{d_{out}}$ represents the local semantic feature learned from the graph-like AST.

\subsection{Cross-language Program Classification}

After that, we obtain the feature vector which includes the global and local semantic features of the input code, and then we perform a fully connected layer for linear dimensional transformation, and eventually the probability $p_i$ is the output through the SoftMax layer. We use the Cross Entropy \cite{de2005tutorial} as our loss function and adopt Adam optimizer \cite{jiang2009automatic} to minimize it. The loss is calculated as follows:

\begin{gather}
    J = -\sum_{i=1}^{k} y_i log(p_i)
\end{gather}
where $k$ is the number of program categories, and $y$ is the label of different programs (if the category is i, then $y_i=1$,else $y_i=0$), $p_i$ is the output after the SoftMax layer.

\section{Experimental Design}
\label{SEC:EXP}
This section firstly presents four research questions to be investigated, then describes our experimental datasets, compared baselines, experiments settings and common evaluation metrics in the following subsections.

\subsection{Research Questions}
\label{SEC:RQ}
Our motivation is to verify whether adopting the unified vocabulary and the unified AST feature fusion could improve the performance of our approach, and how those two mechanisms affect the performance of UAST. Taking the above concerns into account, we raise the following Research Questions (RQs):

\begin{center}
\fcolorbox{gray!50}{gray!5}
{\parbox{.97\linewidth}
{\textbf{RQ1: }How effective is the proposed UAST compared with the other baselines?}
}
\end{center}
RQ1 is intended to investigate whether the proposed UAST outperforms the other state-of-the-art baselines. There are many other works also explore code representation learning semantics for cross-language program classification, and we select representative ones as baselines and will be described in Section 3.3.

\begin{center}
\fcolorbox{gray!50}{gray!5}
{\parbox{.97\linewidth}
{\textbf{RQ2:} How does the unified vocabulary affect the performance of our proposed UAST ?}
}
\end{center}
RQ2 aims to explore whether unified vocabulary could reduce the difference between different programming languages and thus improve the performance of our proposed method. To analyze the impact of unified vocabulary on model effectiveness, we verify the impact of using the unified vocabulary or not on the overall performance through ablation experiments.

\begin{center}
\fcolorbox{gray!50}{gray!5}
{\parbox{.97\linewidth}
{\textbf{RQ3:} How does the unified AST feature fusion affect the performance of our proposed UAST?}
}
\end{center}
RQ3 is put forward to evaluate the impact of the unified AST feature fusion. Since the two sub-networks (SAST and GAST) mentioned in Section 2.3 and 2.4 could extract global syntactic features and local semantic structure features respectively, so it is a need to explore an effect of the fusion mechanism on the learned features of two sub-networks. Therefore, we conduct ablation experiments to verify the impact of using the unified AST feature fusion mechanism on the overall performance.

\begin{center}
\fcolorbox{gray!50}{gray!5}
{\parbox{.97\linewidth}
{\textbf{RQ4:} How do different parameter settings affect the performance of our proposed UAST?}
}
\end{center}
RQ4 is to examine the impact of UAST's own parameters on performance. UAST contains two important parameters which could affect the model performance substantially. One is the length of the path obtained by pre-order traversal of the AST, the other is the layer of GCN which determines how many layers of neighbors' information can be aggregated in the GAST. We set different parameters to find the most suitable parameters which make the best performance.

\subsection{Datasets}
We use two datasets to evaluate the performance of program classification models. The first dataset inherits from the Bi-TBCNN by Bui et al.~\cite{bui2019bilateral}, since the dataset contains two programming languages (Java and C++), we call this dataset Dataset JC. To evaluate our model performance on more programming languages, we collect a dataset from Leetcode which contains 20000 problem solutions of five different programming languages, and we call this dataset Dataset Leetcode. The following is a detailed introduction to these two datasets.

\begin{itemize}
    \item {\verb|JC|}: This dataset includes 10 different categories of programs crawled by Bui et al.~\cite{bui2019bilateral} from GitHub. It contains 5822 Java files and 7019 C++ files. The code files for each category implement the same function, so they are codes that are semantically similar to each other. The dataset is divided into three parts: training set (3), validation set (1), and testing set (1). The specific information of the dataset split is shown in the Table~\ref{tab:dataset}.
    \item{\verb|Leetcode|}: We crawl 50 different categories of programs from Leetcode, each of which contains 400 solutions of different programming languages, with a total of 20000 files. The codes under the same problem are all semantically similar to each other. The dataset includes five different programming languages: C, C++, Java, Python, and JavaScript. We use NICAD\footnote{http://www.txl.ca/txl-nicaddownload.html} to filter out the duplicate code. Considering that each code file contains some salient information (the function name of the solution to each problem written in different programming language is the same, for example, the function name of codes to the problem of "finding the median of two positive-order arrays" are all called \textit{“findMedianSortedArrays”} ), so we use \textit{"XXX"} instead of all function names to each problem and the user-defined function names will not be replaced. Besides, our dataset can also be used as a benchmark dataset for cross-language program classification task. Since the number of problem solutions in different programming languages is unbalanced, we divide it according to the number of different programming languages, and the ratio of division is: training(3): validation(1): testing(1). The specific division information of the dataset is shown in the Table~\ref{tab:dataset}.
\end{itemize}

\begin{table}[h]
\small
% \footnotesize
\caption{The division of Dataset JC and Leetcode.}
\label{tab:dataset}
\begin{tabular}{cccllccccc}
% \hline
\toprule
 & \multicolumn{4}{c|}{JC}            & \multicolumn{5}{c}{Leetcode}            \\ \cline{2-10} 
                      & Java    & \multicolumn{3}{c|}{C++} & C   & C++  & Java & Python & JavaScript \\ \hline
Training               & 3498    & \multicolumn{3}{c}{4215} & 331 & 3428 & 5051 & 2633   & 557        \\
Validation             & 1162       & \multicolumn{3}{c}{1402}    & 110 & 1143 & 1684 & 878    & 185        \\
Testing                & 1162     & \multicolumn{3}{c}{1402}  & 110 & 1143 & 1684 & 878    & 185        \\ 
\bottomrule
% \hline
\end{tabular}
\end{table}

\subsection{Baselines}
We compare UAST with the following state-of-the-art baselines for the cross-language code learning.

\begin{itemize}
    % \item \texttt{Bi-TBCNN}~\cite{bui2019bilateral}: It is an algorithm that uses tree neural networks to classify cross-language programs. The algorithm proposes a bi-NN framework, which builds one neural network on the basis of two sub-networks. Each sub-network encodes a language code grammatically and semantically to achieve cross-language program classification.
    \item \texttt{CodeBERT}~\cite{feng2020codebert}: It is a variant of BERT \cite{devlin2019bert} and it uses RTD (Replaced Token Detection) for pre-training and adopts deep bi-transformer components, so it finally generates the code features that can integrate context information, which can effectively extract the features of the input sequence. And it learns a pre-trained model for 6 programming languages, so it can be used for cross-language program classification.
    \item \texttt{Infercode}~\cite{bui2021infercode}: This method applies the self-supervised learning ideas in natural language processing to the AST of the code, and trains the code representation by predicting the automatic context sub-trees of the AST. Because it has been trained on multiple programming languages, it can extract the characteristics of the different programming languages and can be used to classify cross-language programs.
\end{itemize}

\subsection{Experimental Setting}
We implement our network in Pytorch. For hardware devices, all experiments are run on a 10-core 3.70GHz Intel(R) Core(TM) i9-10900X CPU and NVIDIA GeForce RTX 3090 GPU server. To ensure the fairness of the comparative experiments, we use the same training set for training. We tune the parameters on the validation set and the specific parameter settings are as follows: the epoch of UAST is 5, the batch size is 64, we adopt Adam~\cite{kingma2015adam} as the training optimizer, and the learning rate is set to 0.001. As for SAST, the length of the path sequence is unified to 700 for Dataset JC and 200 for Dataset Leetcode (specifically, if it is insufficient, it is padded with 0; if it is exceeded, it is directly truncated). Besides, the embedding dimension of the path is 200, the number of units of the Bi-LSTM is 64, the number of layers of Bi-LSTM is 2, and the dropout rate of Bi-LSTM is set to 0.5; the dropout rate of self-attention is 0.2, and the number of heads of the self-attention layer is 4. As for GAST, the size of the adjacency matrix of the graph is unified to [400, 400], and the number of layers of GCN is 2.

\subsection{Evaluation Metrics}
For multi-classification tasks, it is common to use public evaluation indicators (Precision, Recall, F1-score, Accuracy) to evaluate the model performance \cite{zhang2019novel}. In order to verify the effect of our model, we use those four evaluation metrics. Detailed definition of those metrics as follows:

% \begin{itemize}
% \item {\verb|Precision|}: Precision indicates the proportion of all predicted positive samples that are actually positive samples for a specific class.
\begin{gather}
    Precision = \frac{\overline{TP}}{\overline{TP}+\overline{FP}}\\
    Recall = \frac{\overline{TP}}{\overline{TP}+\overline{FN}}\\
    F1-score = 2 \times \frac{Precision \times Recall}{Precision + Recall}\\
    Accuracy = \frac{\overline{TP}+\overline{TN}}{\overline {TP}+\overline {FP}+\overline{TN}+\overline{FN}}
\end{gather}

% \end{itemize}
where $\overline{TP}$ (True Positive) is the weighted average of the number of samples that are correctly predicted as positive examples for each class; $\overline{FN}$ (False Negative) is the weighted average of the number of samples that are incorrectly predicted as negative examples per class; $\overline{FP}$ (False Positive) is the weighted average of the number of samples that are incorrectly predicted as positive examples per class; $\overline{TN}$ (True Negative) is the weighted average of the number of samples that are correctly predicted as negative examples per class.

\section{Results and Analysis}
\label{SEC:RES}
This section reports the experimental results by addressing the four research questions that are proposed in Section~\ref{SEC:RQ}.

\subsection{RQ1: Model Performance}
In order to answer RQ1, we compare UAST with mentioned algorithms in Section 3.3. Table \ref{tab:JC} and Table \ref{tab:Leetcode} report the results on two datasets. It can be found from Table \ref{tab:JC} that UAST achieves a Recall of 0.9611, a Precision of 0.9631, a F1-score of 0.9617 and an Accuracy of 0.9626, which outperforms other baselines. Meanwhile, the result on Dataset Leetcode also outperforms other baselines, indicating that our UAST is effective for cross-language program classification.

\begin{table}[h]
\caption{Comparative experiment on Dataset JC.}
\label{tab:JC}
\begin{tabular}{ccccc}
% \hline
\toprule
Model     & Recall          & Precision       & F1-score        & Accuracy \\ \hline
CodeBERT  & 0.9078       & 0.9177            & 0.9090       & 0.9005         \\
Infercode & 0.8317       & 0.8468            & 0.8325       & 0.8343           \\
\textbf{UAST}   & \textbf{0.9611} & \textbf{0.9631} & \textbf{0.9617} & \textbf{0.9626}   \\ 
% \hline
\bottomrule
\end{tabular}
\end{table}

\begin{table}[h]
\caption{Comparative experiment on Dataset Leetcode.}
\label{tab:Leetcode}
\begin{tabular}{ccccc}
% \hline
\toprule
Model     & Recall          & Precision       & F1-score        & Accuracy \\ \hline
CodeBERT  & 0.6147          & 0.6348         & 0.6174      &  0.6245        \\
Infercode &  0.5696               & 0.5819          & 0.5762          & 0.5807         \\
\textbf{UAST}   & \textbf{0.7958} & \textbf{0.8025} & \textbf{0.7965} & \textbf{0.7964}   \\ 
% \hline
\bottomrule
\vspace{-0.6cm}
\end{tabular}
\end{table}

Among them, Infercode learns the characteristics of the sub-tree of the AST through self-supervision and it uses a cross-language corpus for training, so it can be used for cross-language programming classification. However, the extracted feature vector is only 100 dimensions, which cannot fully reflect the characteristics of some long codes, making the final performance worse than the other neural network models. As a powerful pre-trained model, CodeBERT's bi-transformer structure can effectively extract the context information of the code, so its performance can reach 91\% on Dataset JC. However, it learns the feature of code by token embedding, so it cannot reflect the semantic characteristics of the code. UAST comprehensively considers global and local code structure semantic information, and also the unified vocabulary in it can reduce the difference in different programming languages, the performance is better than other baselines.

Furthermore, we conduct experiments on two datasets of different sizes and language types. And the performance of our proposed UAST in both datasets outperforms other baselines, which not only shows the efficiency of our model but also shows that our proposed model has a strong generalization ability.

\begin{center}
\fcolorbox{black}{white}
{\parbox{.97\linewidth}
{\textbf{Result 1:} \textit{Our proposed UAST significantly outperforms other state-of-the-art baselines in terms of Precision, Recall, F1-score, and Accuracy.}}
}
\end{center}

\subsection{RQ2: The Impact of the Unified Vocabulary}
As mentioned in Section 2.2, we propose the unified vocabulary mechanism to reduce the differences between different programming languages. So in order to answer RQ2, we conduct an ablation experiment to explore the impact of the unified vocabulary on model performance. The results are shown in the Table \ref{tab:ablation_uv}.\footnote{Method with -V indicates that the unified vocabulary is not used.}

According to Table \ref{tab:ablation_uv}, we find that using the unified vocabulary outperforms those who do not use it by 0.37\% - 3.32\% in terms of Recall, by 0.37\% - 3.86\% in terms of Precision, by 0.42\% - 3.26\% in terms of F1-score, and by 0.30\% - 2.86\% in terms of Accuracy on both Dataset JC and Leetcode, which indicates that the unified vocabulary indeed reduces the feature gap between different programming languages and thus improves the performance of cross-language program classification. 

\begin{table}[h]
\caption{Ablation experiment results of the unified vocabulary on Dataset JC and Leetcode.}
\begin{tabular}{cccccc}
% \hline
\toprule
Dataset                   & Model                                 & Recall          & Precision       & F1-score             & Accuracy        \\ \hline
\multirow{6}{1cm}{JC} 
           & \multicolumn{1}{c|}{SAST-V}             & 0.8802          & 0.8868          & 0.8816          & 0.8856          \\
        & \multicolumn{1}{c|}{\textbf{SAST}} & \textbf{0.9125} & \textbf{0.9254} & \textbf{0.9142} & \textbf{0.9142} \\
            & \multicolumn{1}{c|}{GAST-V}             & 0.9467          & 0.9479          & 0.9469          & 0.9478          \\
                          & \multicolumn{1}{c|}{\textbf{GAST}} & \textbf{0.9504} & \textbf{0.9516} & \textbf{0.9511} & \textbf{0.9508} \\
                          & \multicolumn{1}{c|}{UAST-V}             & 0.9524          & 0.9526          & 0.9509          & 0.9516          \\
                          & \multicolumn{1}{c|}{\textbf{UAST}} & \textbf{0.9611} & \textbf{0.9631} & \textbf{0.9617} & \textbf{0.9626} \\ \hline
\multirow{6}{1cm}{Leetcode} 
                         & \multicolumn{1}{c|}{SAST-V}             & 0.6553          & 0.6894          & 0.6540          & 0.6554          \\
                          & \multicolumn{1}{c|}{\textbf{SAST}} & \textbf{0.6718} & \textbf{0.7020} & \textbf{0.6707} & \textbf{0.6721} \\
      & \multicolumn{1}{c|}{GAST-V}             & 0.7744          & 0.7793          & 0.7749          & 0.7749          \\
                          & \multicolumn{1}{c|}{\textbf{GAST}} & \textbf{0.7892} & \textbf{0.7956} & \textbf{0.7887} & \textbf{0.7892} \\
                          & \multicolumn{1}{c|}{UAST-V}             & 0.7893          & 0.7970          & 0.7904          & 0.7882          \\
                          & \multicolumn{1}{c|}{\textbf{UAST}} & \textbf{0.7958} & \textbf{0.8025} & \textbf{0.7965} & \textbf{0.7964} \\ 
\bottomrule
\label{tab:ablation_uv}
\vspace{-0.6cm}
\end{tabular}
\end{table}

% The following analysis provides insight into why dictionaries can be beneficial for performance:
Since different programming languages have their own unique coding features, the terms obtained by the parser in parsing different programming languages are different even if they have the same meaning, so the AST nodes will show different node names. If we directly generate a vocabulary for all the node names of the AST, then each language will have a certain difference in its corresponding AST, which will enlarge the gap of code features. So the use of the unified vocabulary will reduce the difference in terms obtained by code parsing, and further reduce the feature learned by the proposed network, so the performance for cross-language program classification tasks has been improved.

Furthermore, we find that in addition to improving the performance of all three networks (SAST, GAST, and UAST), the use of the unified vocabulary in the SAST network has the most pronounced effect. Due to the fact that the SAST network is designed to extract the features of a traversal path sequence of the AST, and the initial embedded vector input into the network is determined by vocabulary mapping, so the unified vocabulary has the most direct impact on the SAST network which leads to the notable performance. As for the GAST network, the adjacency matrix is not constructed based on the unified vocabulary, but only on the structure of the AST. And the feature matrix is composed of the one-hot vector of the node name, which is related to the unified vocabulary, so the overall impact of the unified vocabulary on the GAST is relatively insignificant compared to SAST.

Given all those results discussed above, the unified vocabulary improves the performance to a certain extent, confirming the effectiveness of the unified vocabulary for the cross-language program classification.

\begin{center}
\fcolorbox{black}{white}
{\parbox{.97\linewidth}
{\textbf{Result 2:} \textit{The unified vocabulary does improve model performance to a certain extent, so leveraging the unified vocabulary to generate the embedded vector is a good choice for cross-language classification tasks.}}
}
\end{center}

\subsection{RQ3: The Impact of the Unified AST Feature Fusion}

Table \ref{tab:ablation_fusion} shows that the performance after the unified AST feature fusion is 1.06\% - 4.86\% higher than that before fusion on Dataset JC, and 0.66\% - 12.58\% higher than that before fusion on Dataset Leetcode, indicating that the unified AST feature fusion could further enhance the effectiveness of our proposed model.

In addition, we find that the performance of SAST is not good as the performance of GAST on both datasets. This result implies that the feature extracted from the flattened AST sequence structure is less useful than the feature extracted from the AST graph-like structure. The main reason is that the sequence obtained by traversing the AST has fewer structural characteristics of the branch statements, cause it only includes the overall process of the source code. While the adjacency matrix in the graph-like AST can capture the associated relationship with neighbor nodes, which is very effective for some structures (i.e., for-loop structure, if-condition structure, etc.), the performance of SAST is not as good as that of GAST.

\begin{table}[h]
% \begin{threeparttable}
\caption{Ablation experiment results of the unified AST feature fusion on Dataset JC and Leetcode.}
\begin{tabular}{cccccc}
% \hline
\toprule
% \multicolumn{2}{c}{}      
Dataset                   & Model  
& Recall & Precision & F1-score    & Accuracy \\ \hline
\multirow{3}{1cm}{JC}                                & \multicolumn{1}{c|}{SAST} & 0.9125 & 0.9254    & 0.9142 & 0.9142   \\
& \multicolumn{1}{c|}{GAST} & 0.9504 & 0.9516    & 0.9511 & 0.9508   \\
                             & \multicolumn{1}{c|}{\textbf{UAST}} & \textbf{0.9611} & \textbf{0.9631} & \textbf{0.9617} & \textbf{0.9626} \\ \hline
 \multirow{3}{1cm}{Leetcode}                        
                            & \multicolumn{1}{c|}{SAST} & 0.6718 & 0.7020    & 0.6707 & 0.6721   \\
       & \multicolumn{1}{c|}{GAST} & 0.7892 & 0.7956    & 0.7887 & 0.7892   \\
                            & \multicolumn{1}{c|}{\textbf{UAST}} & \textbf{0.7958} & \textbf{0.8025} & \textbf{0.7965} & \textbf{0.7964} \\
                            
% \hline
\bottomrule
\label{tab:ablation_fusion}
\vspace{-0.4cm}
\end{tabular}
\end{table}
% \vspace{0.5cm}

In order to further analyze the reasons why unified AST feature fusion works better, we perform a reasonable analysis as follows. The two code features extracted by the unified AST neural network (SAST and GAST) contain global features and local features of the code respectively. If we only consider one of the features, then the features of the other aspect will be ignored, so fusing the two hidden vectors can strengthen the features of the corresponding dimensions, which results in better performance. Furthermore, the SAST focuses on global information extraction, and the GAST pays attention to the capture of local information. Therefore, two code features are spliced and fused after the unified AST neural network (SAST, GAST), which can comprehensively consider global and local information and learn the semantic structure characteristics of the code from multiple aspects.

Given all those factors stated above, the unified AST feature fusion mechanism can indeed improve the performance of the network as a whole.

\begin{center}
\fcolorbox{black}{white}
{\parbox{.97\linewidth}
{\textbf{Result 3:} \textit{The unified AST feature fusion mechanism leverages both global structural information and local semantic information for code representation, which can significantly help improve model performance.}}
}
\end{center}

\subsection{RQ4: The Impact of Parameter Settings}
\begin{figure}[h]
    \centering
    \subfigure[Results on Dataset JC]{
        \includegraphics[width=0.47\linewidth]{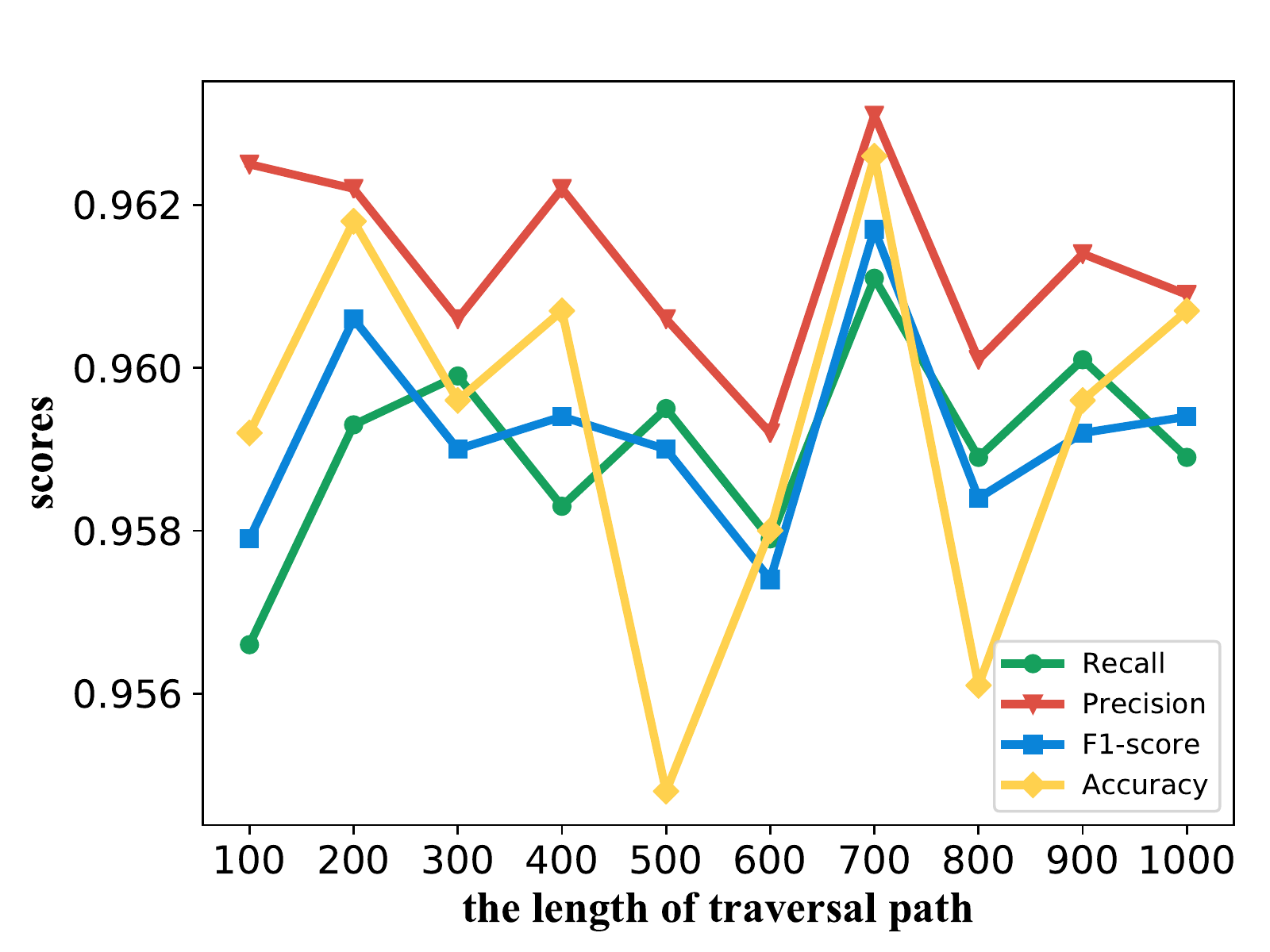}
        }
    \subfigure[Results on Dataset Leetcode]{
        \includegraphics[width=0.47\linewidth]{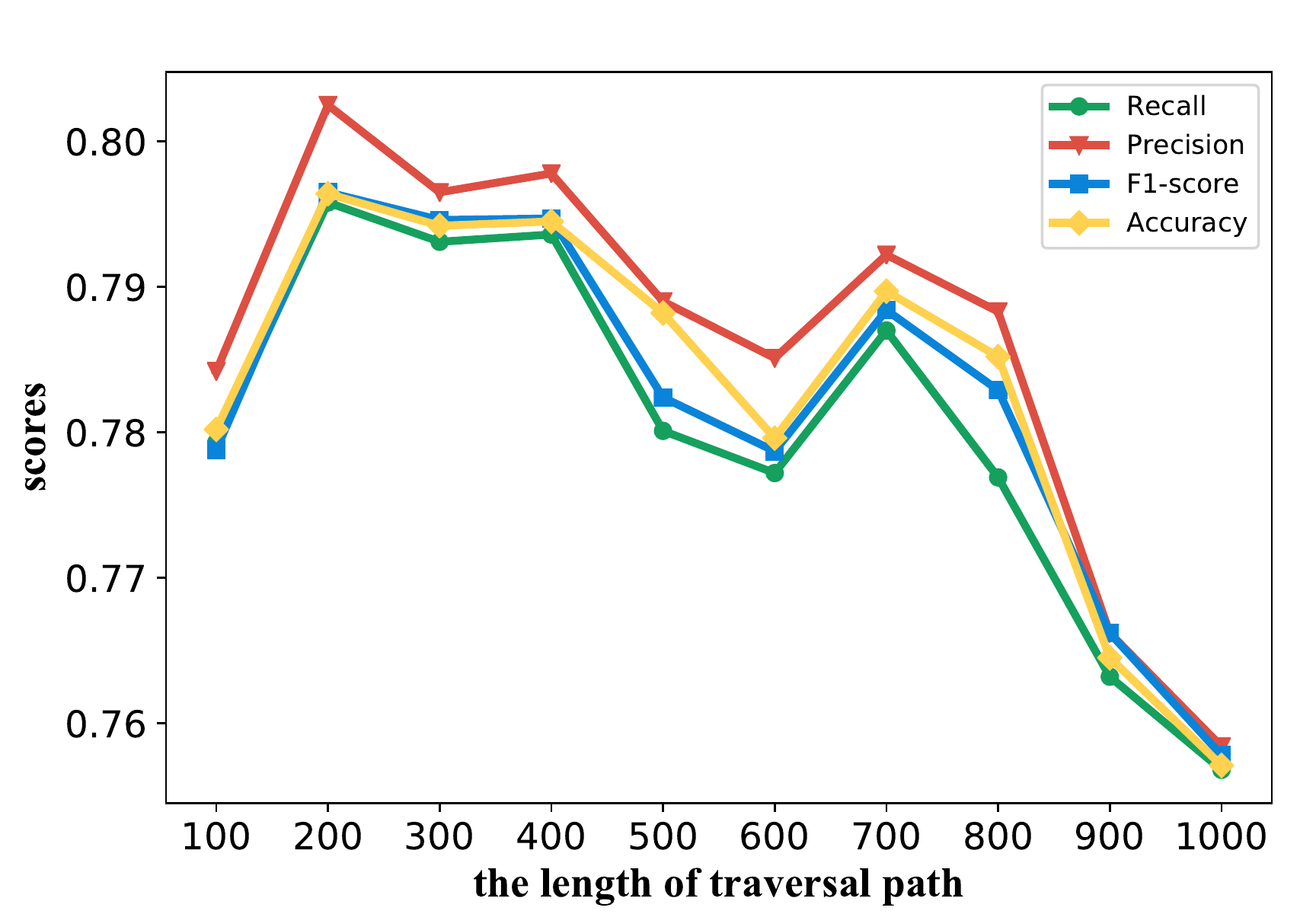}
        }
    \caption{The effect of the length of the path sequence in SAST network on model performance. }
    \label{fig:length}
\end{figure}

\begin{figure}[h]
    \centering
    \subfigure[Results on Dataset JC]{
        \includegraphics[width=0.47\linewidth]{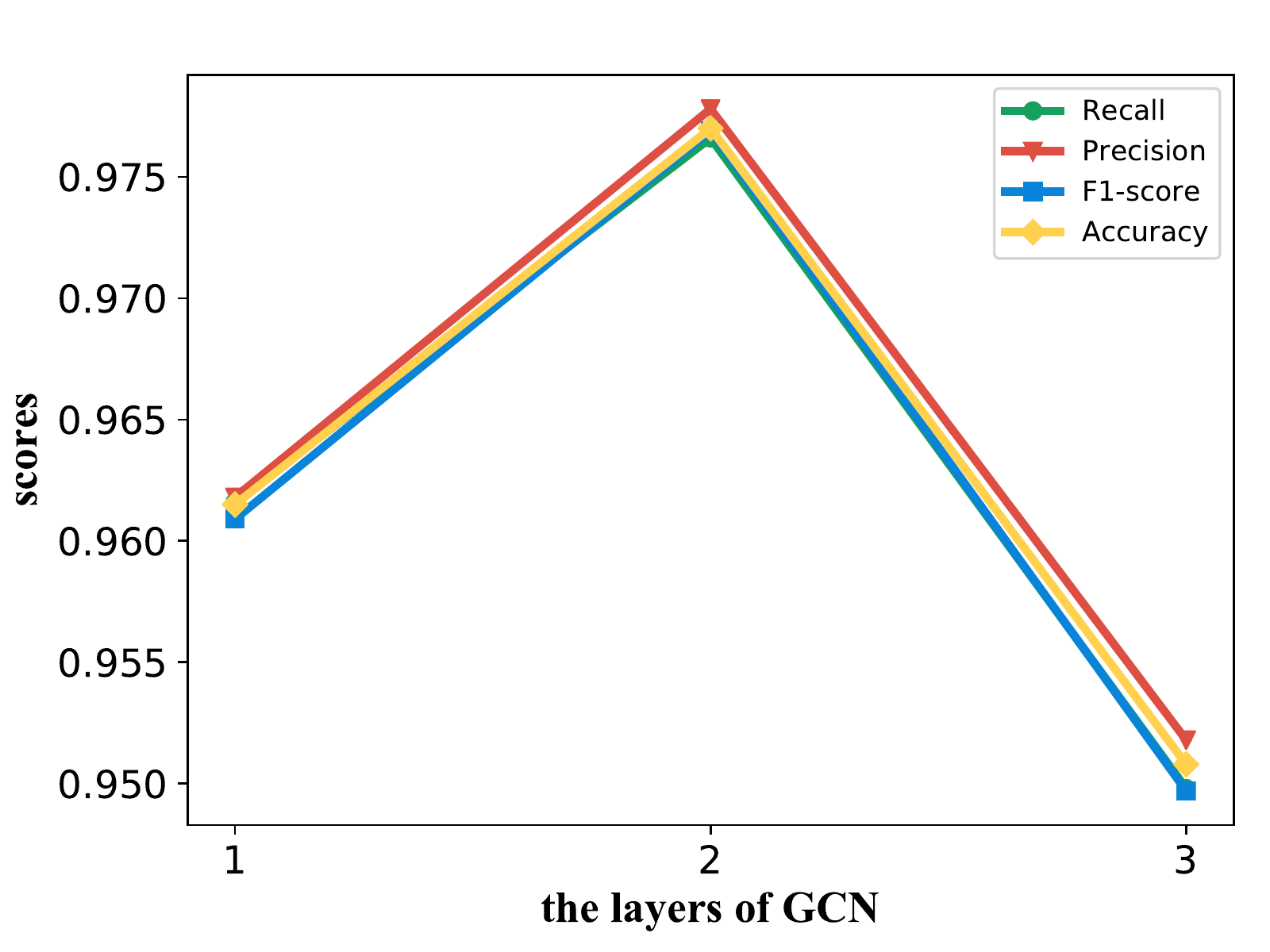}
        }
    \subfigure[Results on Dataset Leetcode]{
        \includegraphics[width=0.47\linewidth]{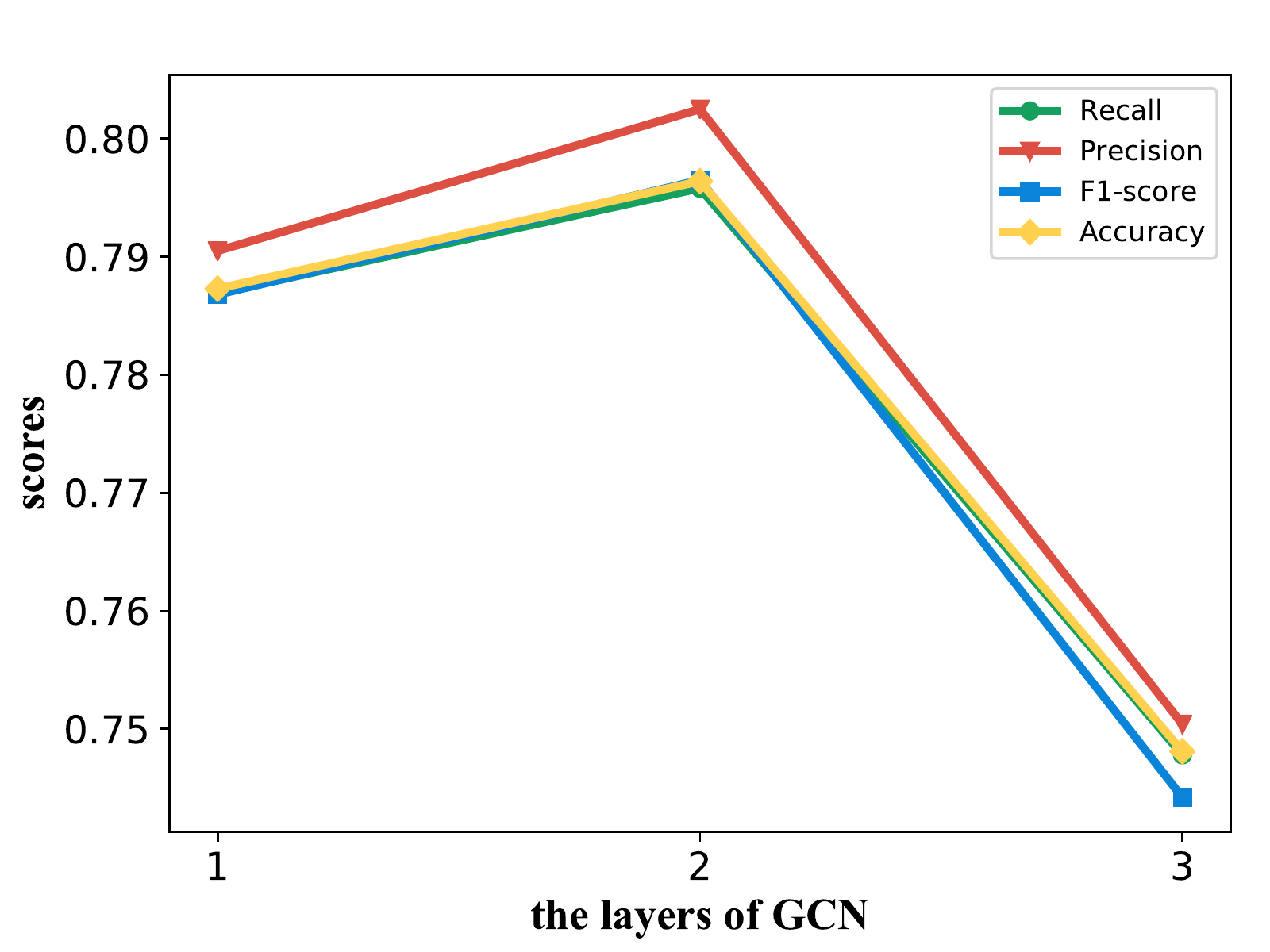}
        }
    \caption{The effect of the layer of GCN in GAST Network on model performance. }
    % \vspace*{-2.0ex}
    \label{fig:GCN}
\vspace{-0.6cm}
\end{figure}

In our proposed UAST, the length of the path sequence in SAST and the layers of GCN in GAST are two important parameters that could affect the performance of the model. The length of the path sequence determines how much code content is fed into the neural network. Due to the unbalanced distribution of the length of the source code in the dataset, setting the appropriate length is of particular importance. The number of layers of GCN in GAST, namely hop, represents the range that each node can aggregate information from neighbor nodes. As the number of hops increases, each node can aggregate a larger range of information from neighboring nodes, thereby focusing on a larger range of local semantic information of the code. To answer RQ4, we perform experiments with different parameter settings, and the experimental results are shown in Figure \ref{fig:length} and Figure \ref{fig:GCN}.

% \begin{figure}
% \centering
% \includegraphics[width=\linewidth]{leetcode.pdf}
% \caption{The Effect of the Length of the Path Sequence in SAST Network on Model Performance.}
% \label{fig:leetcode_len}
% \end{figure}

% \begin{figure}
% \centering
% \includegraphics[width=\linewidth]{JC.pdf}
% \caption{The Effect of the Length of the Path Sequence in SAST Network on Model Performance.}
% \label{fig:JC_len}
% \end{figure}

% \begin{figure*}[]
%     \centering
%     \subfigure[JC]{
%         \includegraphics[width=0.4\linewidth]{JC.pdf}
%         }
%     \subfigure[Leetcode]{
%         \includegraphics[width=0.4\linewidth]{leetcode.pdf}
%         }
%     \caption{ Demo }
%     \label{fig}
% \end{figure*}

% \begin{figure}[]
%     \centering
%     \subfigure[Results on Dataset JC]{
%         \includegraphics[width=\linewidth]{JC.pdf}
%         }
%     \quad
%     \subfigure[Results on Dataset Leetcode]{
%         \includegraphics[width=\linewidth]{leetcode.pdf}
%         }
%     \caption{The Effect of the Length of the Path Sequence in SAST Network on Model Performance. }
%     \label{fig:length}
% \end{figure}

For the length of the path sequence, we explore the effect of different lengths on performance by setting different length values (100-1000). Figure \ref{fig:length} shows that the UAST achieves the best performance when the length is set to 700 on Dataset JC and 200 on Dataset Leetcode in terms of Recall, Precision, F1-score and Accuracy. As the length increases, the effect of the model will not be improved, but it will increase the complexity of the model. Besides, Bi-LSTM cannot effectively handle long sequences of inputs, which is one of the possible reasons for the poor performance. When the length is reduced, the performance of the model will also decrease. The main reason is that reducing the length of the input will cut out some useful information, so that the neural network cannot effectively capture some key feature of the code. 

Notably, we have statistics on the length of AST path for both datasets. As shown in the Table \ref{tab:statistic}, we find that 80\% of the AST path sequences are within 726 in length on the Dataset JC, while 80\% of the AST path sequences are within 221 in length on the Dataset Leetcode. Therefore, it is most suitable to set the sequence input to 700 and 200 respectively on Dataset JC and Leetcode, because long input will cause some short codes to be filled with 0, and short input will lose some key information. Thus, when training on different datasets, it is a need to have statistics on the length of AST path and choose the most appropriate length before training.

\begin{table}[h]
\caption{Statistical distribution of AST path sequence length on Dataset JC and Leetcode.}
\begin{tabular}{cccccc}
\hline
Dataset  & Mean & Median & 70\% & 80\% & 90\% \\ \hline
JC       & 576  & 354    & 502  & 726  & 1498 \\
Leetcode & 165  & 144    & 189  & 221  & 279  \\ \hline
\label{tab:statistic}
\end{tabular}
\end{table}

% It is worth noting that we have made statistics on the code AST path lengths of the two datasets. As shown in Table \ref{fig:statistics}, we found that the median length of the AST path sequences generated by the codes in the two datasets is about 400 Therefore, choosing 400 is also the most suitable, because too long will cause some short codes to be completed as 0, and too short will lose some key information. Therefore, when training on different datasets, it is more reasonable to first check the median length of the path sequence in the dataset before proceeding.

For the layers of GCN, we explore the effect of different layers on performance by setting different layers of GCN (1,2,3). It can be seen from Table \ref{fig:GCN} that the model performs best when the number of layers of GCN is set to 2. The first-order neighbors indicate a smaller neighbor range, resulting in a smaller captured tree structure, so the performance is not as good as two GCN layers stacking. A larger number of layers represents a wider range of neighbors, and the experimental results are even worse than that of one layer. Our initial inference is that some specific structures are only displayed in the range of two layers when writing the code. In addition, the increase in the number of layers could make the model more complex and lead to overfitting.

\begin{center}
\fcolorbox{black}{white}
{\parbox{.97\linewidth}
{\textbf{Result 4:} \textit{For the choice of path length, it is better to have statistics on dataset AST path before training and choose the most suitable value. While setting the layer of GCN to 2 is beneficial to model effectiveness.}}
}
\end{center}

% \begin{figure*}[]
%     \centering
%     \subfigure[Results on Dataset JC with various path length in SAST]{
%         \includegraphics[width=0.48\linewidth]{FIG/JC.pdf}
%         }
%     \subfigure[Results on Dataset Leetcode with various path length in SAST]{
%         \includegraphics[width=0.48\linewidth]{FIG/leetcode.pdf}
%         }
%     \quad
%     \subfigure[Results on Dataset JC with various number of GCN layers in GAST]{
%         \includegraphics[width=0.48\linewidth]{FIG/GCN_JC.pdf}
%         }
%     \subfigure[Results on Dataset Leetcode with various number of GCN layers in GAST]{
%         \includegraphics[width=0.48\linewidth]{FIG/GCN_Leetcode.pdf}
%         }

%     \caption{The Effect of the Length of the Path Sequence in SAST Network and the Layer of GCN in GAST Network on Model Performance. }
%     \label{fig:length}
% \end{figure*}

% \begin{figure}[htbp]
% \centering
% \begin{minipage}[t]{0.48\textwidth}
% \centering
% \includegraphics[width=\linewidth]{JC.pdf}
% \caption{World Map}
% \end{minipage}
% \begin{minipage}[t]{0.48\textwidth}
% \centering
% \includegraphics[width=\linewidth]{JC.pdf}
% \caption{Concrete and Constructions}
% \end{minipage}
% \end{figure}
% 。

\section{Discussion}
\label{SEC:DIS}
This section discusses the considerations behind the proposed approach and the threats to validity of this study.

\subsection{Why we choose AST as the code representation form?}
In order to capture the characteristics of the code, current research is mainly to learn the code by exploring different code  \textit{Intermediate Representations (IRs)}. \textit{Token} represents code lexical information, \textit{Abstract Syntax Tree (AST)} is a representation form that includes code structure and grammatical information. Graph-based IRs such as \textit{Control Flow Graph (CFG)} and \textit{Data Flow Graph (DFG)} contain the control flow and data flow characteristics of the code. Different representation forms include different features from different views of the code. In recent years, many excellent studies based on those IRs have been carried out~\cite{azcona2019user2code2vec, mou2016convolutional, feng2020codebert, ben2018neural}. It is currently the mainstream method of program classification to distinguish codes of different categories (functions) by capturing code characteristics.

% However, most of the previous researches only focused on single language program classification, and few researches conducted on cross-language program classification tasks.

For the intermediate representation of the code, we have a variety of options, token, AST or graph-based forms (such as CFG, DFG). For the token, it only contains the information of the lexical level of the code, and the program classification is based on the function implemented by the code (i.e., the semantic level), so the token is not the best choice. As for graph-based representations, although some control information they contain can describe the semantics of the code, it is not easy to generate the graph of multiple languages, and the cost of training a pure graph model is high. While AST is relatively easy to generate for most programming languages. And as a tree structure, AST is easier to traverse. So we finally decide to use AST as our code representation. 

\subsection{Why does UAST work?}
The essential difficulty of cross-language program classification is to learn the semantic features of codes implemented in different programming languages. This model achieves such excellent performance on cross-language program classification mainly due to two mechanisms (unified vocabulary and unified AST feature fusion) in it. The unified vocabulary reduces the differences between different programming languages, making it more effective for cross-language classification. The fusion mechanism fuses the learned code features of global and local information to capture the semantics of the code leading the classification performance better.

\subsection{Threats to Validity}
There are two types of threats to this study, and accordingly, we have made efforts to mitigate those threats to validity.

% 因变量只受(唯一的)自变量影响
\emph{Internal threat: } The threats to the internal validity of this study may result from the dataset. Since we collect the data from Leetcode, where the function names of the solutions to the problems are the same. If they explicitly appear in the code, the label will be exposed. So we substitute all the function name of solutions to eliminate this threat. Meanwhile, the dataset may be some duplicate codes, therefore, we deduplicate the similar codes using NICAD tools to reduce this threat.

% Another related threat is that some noise data (e.g., % casual/superficial comments such as “OK”, “fine”, etc.) exists in % both training set and test set, which may not be able to guaran-
% tee a qualified reviewer recommendation. However, the evaluation
% between HGRec and other recommenders is based on the same
% dataset, which may mitigate this threat to a fair degree. 

% 结果、结论需要具备泛化性能
\emph{External threat: } In order to rule out that our proposed method is only effective on our Dataset Leetcode, we use two datasets for experiments, which could reduce the threat of accidental results to a certain extent. The results outperform other baselines using the same two datasets, illustrating our model has good generalization ability.

% 你想度量的东西，是用正确的度量指标/方法在度量，比如希望度量召回率，正确的是用recall，如果用的是其他的，Construct threat就高
% \emph{Construct threat} refers to how well a test measures what it is supposed to measure. In this study,

% \begin{table*}[]
% \begin{tabular}{lllllllll}
% \hline
%       & \multicolumn{4}{c}{Java/CPP}       & \multicolumn{4}{c}{leetcode}        \\ \cline{2-9} 
% model & recall & precision & F1 & accuracy & recall & precision & F1  & accuracy \\ \hline
% AST&0.8812&0.8854&0.8826&0.8816&0.8812&0.8854&0.8826&0.881 \\
% AST   &0.8812&0.8854&0.8826&0.8816&0.8812&0.8854&0.8826&0.881 \\ \hline
% \end{tabular}
% \end{table*}

\section{Related Work}
\label{SEC:REL}

\subsection{Program Classification}
Program classification can be regarded as a high-level abstraction of code. In all source code mining tasks, program classification lays the foundation for various tasks related to source code understanding. Program classification can be applied to many scenarios in the field of software engineering, such as code clone detection~\cite{bellon2007comparison, borstler1995feature}, bug fixing \cite{peters2017text}, code smell classification \cite{fontana2017code}, defect classification \cite{wang2016automatically}, program understanding \cite{alon2019code2vec}, etc. The earliest research on program classification dates back to the last century. \cite{clark1980algorithm, borstler1995feature, jiang2009automatic, taherkhani2011recognizing} used features (such as token, component name, counts of statements, code metrics, etc) to classify programs. Since these studies are classified based on artificially defined rules and extract features at the surface level of the code, their performances are not too prominent, but it starts the beginning of exploration for program classification.

Machine learning has been shown to yield promising results for classification. Ugurel et al.~\cite{ugurel2002s} used Support Vector Machine (SVM) \cite{drucker1997support} to classify the code's token set, and Ma et al.~\cite{ma2018automatic} used SVM, Decision Tree \cite{quinlan1986induction}, and Bayesian Network \cite{friedman1997bayesian} to classify the code based on the code's token sequence to determine which types software artifacts are produced by various open source projects at different levels of granularity. Shimonaka et al.~\cite{shimonaka2016identifying} used four machine algorithms to construct a learning model for the syntactic information of the code, which is used to identify the source code to determine whether the code is automatically generated code.

With the rising trend of deep learning, more and more researches begin to combine neural networks to learn the characteristics of the code. Mou et al.~\cite{mou2016convolutional} proposed a tree-based convolutional network, the kernel of which can be used to capture the structural information of the source code, and performed well in the task of classifying programs. Zhang et al.~\cite{zhang2019novel} divided the AST into small sub-ASTs, encoded those sub-ASTs into vectors by capturing the lexical and syntactic features, and then used the Bi-RNN model to generate the code vector representation. This algorithm has achieved excellent results in the source code classification task. Barchi et al.~\cite{barchi2021exploration} explored the use of Convolutional Neural Networks (CNN) \cite{kalchbrenner2014convolutional} to analyze program source code and proved that the CNN model can be successfully applied to source code classification. Compared with the most advanced methods, this method provided higher accuracy and less learning time.

% \vspace{-0.5cm}

\subsection{Code Representation Learning}
The emergence of deep neural networks technology has brought new solutions to the field of software engineering, and at the same time, an increasing amount of attention has been devoted to the learning of code representation. Code representation forms are used to analyze the original meaning of the source code at different levels. For example, the token represents the lexical information of the code, while the Abstract Syntax Tree (AST) represents the code structure and grammatical information obtained by the syntax analyzer, and the Data Control Flow (DFG) and the Control Flow Graph (CFG) represent the code data flow and control flow information. Also, there are other intermediate representations (IRs) to represent the code, and different representation forms combined with neural network to learn code semantics is the direction that researchers are exploring currently. 

Harer et al.~\cite{harer2018automated} used Word2Vec \cite{mikolov2013efficient} tool to generate the initial embedded vector for the C/C++ token, and then used the TextCNN \cite{zhang2015sensitivity} model to learn the features of the vector for software vulnerability detection. Azcona et al.~\cite{azcona2019user2code2vec} analyzed the Python code submitted by students by embedding the token into a vector, and then extracted the effective features of the student to analyze the progress and performance of the student. The token contains the lexical information of the code, but ignores the structural features and grammatical logic information of the code.

Mou et al.~\cite{mou2016convolutional} proposed a convolutional neural network named TBCNN based on AST. They designed a sub-tree kernel to slide on the AST to extract the structure information of the tree, and also used dynamic pooling to deal with the number of sub-trees of the AST. Uri Alon et al.~\cite{alon2019code2vec} proposed a code embedding algorithm called code2vec, which gives an attention score to each path in the AST, so that it could extract the syntactic information of the entire source code while dividing sentences of different degrees of importance. Wang et al.~\cite{wang2020detecting} constructed a graph called \textit{Flow-Augmented Abstract Syntax Tree} (FA-AST) to represent source code, which contains control flow and data flow information. Then they used Graph Neural Networks (GNN)~\cite{scarselli2008graph} to capture the feature of FA-AST.

Ben-Nun et al.~\cite{ben2018neural} tried to learn code semantics based on the IR of the code. They converted IR into \textit{context flow graph} (XFG), which contains the data flow and control flow of the code, and then used neural networks to learn the code features from XFG embedding. Wei et al.~\cite{wei2020lambdanet} generated a \textit{type dependency graph} from source code, which links type variables with logical constraints as well as name and usage information, and then used GNN to extract the feature of this kind of IR.

\subsection{Cross-language Code Learning}
% It is common that one program is composed of multiple programming languages, and therefore we should concern the cross-programming language problem.

Cross-language learning is widely used in machine translation in the field of NLP \cite{huang2013cross, conneau2019cross}, and cross-language code learning is still in its infancy. Nguyen et al.~\cite{nguyen2014migrating} proposed a tool called semSMT based on SMT to migrate Java programs to C\#, and it operated at all three lexical, syntactic, and semantic levels to extract code features. Bui et al.~\cite{bui2019bilateral} proposed a bilateral dependency neural networks to learn the features of different languages, and then used Siamese network for similarity detection. Ye et al.~\cite{ye2020misim} proposed the MISIM (Machine Inferred Code Similarity System) model, in which the Context-Aware Semantic Structure (CASS) can capture the code's contextual semantics to describe the intent of the code, and CASS could also learn the language-independent representation through manual configuration for cross-language detection. The Infercode proposed by Bui et al.~\cite{bui2021infercode} implemented self-supervised learning by predicting the automatic context sub-trees of the AST, in which vectors are generated by training multiple languages, and it performed well on multiple tasks.The TPTrans proposed by Peng et al.~\cite{peng2021integrating} performed feature learning for different programming languages, and incorporated the feature information of the code tree structure into the transformer structure using position embedding and embedding. This model performed well for code summarization tasks.

\section{Conclusions}
\label{SEC:CON}
In order to better address the problem of cross-language code learning, in this study, we propose a Unified Abstract Syntax Tree neural network (UAST) framework for cross-language program classification task. The UAST contains two sub-networks (SAST and GAST), where SAST is used to extract the global code syntactic features contained in the AST path sequence, GAST is used to capture the local code semantic features of the AST tree. And UAST network can comprehensively consider global and local code features through the unified AST feature fusion, which could effectively learn code semantic features. In addition, the unified vocabulary mechanism we proposed can reduce the difference between different programming languages. The comparative experiments on the public dataset and our collected dataset both show that the UAST performs better than other state-of-the-art baselines by a significant margin, and can better distinguish between different programs. In conclusion, this study leverages the powerful representation learning techniques to model cross-language source code, which contributes to the state-of-the-art AI for software engineering. With regard to the future work, we are going to leverage both source code and natural text information to better improve program classification.

\section*{ACKNOWLEDGEMENTS}
\label{SEC:ACK}
This work is supported by the National Key Research and Development Project (No.2019YFE0105500), the National Natural Science Foundation of China (No.62072227), the Research Council of Norway (No.309494), the Key Research and Development Program of Jiangsu Province (No.BE2021002-2), the Natural Science Foundation of Chongqing (No. cstc2021jcyj-msxmX0538), and the Chongqing Science and Technology Plan Project (No.cstc2018jszx-cyztzxX0037).

\balance
\bibliographystyle{ACM-Reference-Format}
\bibliography{ref}

\end{document}